\documentclass{article}

\usepackage[english]{babel}

\usepackage[letterpaper,top=2cm,bottom=2cm,left=3cm,right=3cm,marginparwidth=1.75cm]{geometry}

\usepackage{amsmath}
\usepackage{amsfonts}
\usepackage{mathtools}
\usepackage{graphicx}
\usepackage[colorlinks=true, allcolors=blue]{hyperref}
\usepackage[disable]{todonotes}
\setuptodonotes{inline}
\usepackage[nameinlink]{cleveref}
\usepackage{siunitx}
\usepackage{glossaries}
\usepackage{authblk}

\newcommand{\CPP}{C\kern-0.05em\raisebox{0.10em}{\scalebox{0.8}{++}}}

\newcommand{\gmax}{g_\mathrm{max}}

\newcommand{\Nnrn}{N_\mathrm{nrn}} %
\newcommand{\SynsPerHalfRow}{S_\mathrm{H,nrn}} %

\newcommand{\SynType}{\textrm{SynType}} %

\newcommand{\Addr}{\textrm{SrcAddr}}
\newcommand{\TgtNrn}{\textrm{TgtNrn}}
\newcommand{\Side}{\textrm{Side}}

\newcommand{\Decoder}{\textrm{Pat}}

\newcommand{\Synapses}{S}
\newcommand{\HalfRows}{H}
\newcommand{\Rows}{R}
\newcommand{\Drivers}{D}

\newacronym{CMOS}{CMOS}{complementary metall-oxide-semiconductor}
\newacronym{HICANN}{HICANN}{High Input Count Analog Neural Network}
\newacronym{HICANN-DLS}{HICANN-DLS}{High Input Count Analog Neural Network - Digital Learning System}
\newacronym{FPGA}{FPGA}{Field Programmable Gate Array}
\newacronym{AdEx}{AdEx}{adaptive-exponential integrate-and-fire}
\newacronym{LIF}{LIF}{Leaky Integrate-and-Fire Model} %
\newacronym{OTA}{OTA}{operational transconductance amplifier}
\newacronym{MOSFET}{MOSFET}{metall oxide semiconductor field effect transistor}
\newacronym{DAC}{DAC}{digital-to-analog converter}
\newacronym{STDP}{STDP}{spike-timing-dependent plasticity}
\newacronym{STP}{STP}{short-term plasticity}
\newacronym{ISI}{ISI}{inter-spike interval}
\newacronym{HTD}{HTD}{hardware time domain}
\newacronym{BTD}{BTD}{biological time domain}
\newacronym{LVDS}{LVDS}{low-voltage differential signaling}
\newacronym{AER}{AER}{adress-event representation}
\newacronym{DLL}{DLL}{delay-locked loop}
\newacronym{BEG}{BEG}{background event generator}
\newacronym{DNC}{DNC}{Digital Network Chip}
\newacronym{PCS}{PCS}{Pulse Communication Subgroup}
\newacronym{PCB}{PCB}{Printed Circuit Board}
\newacronym{ASIC}{ASIC}{application-specific integrated circuit}
\newacronym{DDR}{DDR}{double-data-rate}
\newacronym{VLSI}{VLSI}{very-large-scale integration}
\newacronym{CRC}{CRC}{cyclic redundancy check}
\newacronym{FIFO}{FIFO}{First In First Out}
\newacronym{API}{API}{application programming interface}
\newacronym{ARQ}{ARQ}{Automatic Repeat reQuest}
\newacronym{HALBE}{HALBE}{Hardware Abstraction Layer Backend}
\newacronym{STHAL}{STHAL}{Stateful Hardware Abstraction Layer}
\newacronym{ESS}{ESS}{Executable System Specification}
\newacronym{RTL}{RTL}{Register Transfer Level}
\newacronym{SRAM}{SRAM}{static random-access memory}
\newacronym{MSB}{MSB}{most significant bit}
\newacronym{LSB}{LSB}{least significant bit}
\newacronym{JTAG}{JTAG}{joint test action group}
\newacronym{FCP}{FCP}{TODO}
\newacronym{PSP}{PSP}{postsynaptic potential}
\newacronym{MC}{MC}{minicolumn}
\newacronym{HC}{HC}{hypercolumn}
\newacronym{PYR}{PYR}{pyramidal cells}
\newacronym{RSNP}{RSNP}{regular spiking non-pyramidal cells}
\newacronym{BAS}{BAS}{basket cell}
\newacronym{WTA}{WTA}{winner-take-all}
\newacronym{RS}{RS}{regular spiking cells}
\newacronym{FS}{FS}{fast spiking cells}
\newacronym{AI}{AI}{asynchronous irregular}
\newacronym{FDSOI}{FDSOI}{fully depleted silicon on isolator}
\newacronym{HBP}{HBP}{Human Brain Project}
\newacronym{MLP}{MLP}{multi-layer perceptron}
\newacronym{SNN}{SNN}{spiking neural network}
\newacronym{DNN}{DNN}{deep neural network}
\newacronym{ReLU}{ReLU}{rectified linear unit}
\newacronym{SIMD}{SIMD}{single instruction, multiple data}
\newacronym{ADC}{ADC}{analog-to-digital converter}
\newacronym{QUBO}{QUBO}{quadratic unconstrained binary optimization}
\newglossaryentry{PyNN}{name={PyNN}, description={A Python package for simulator-independent specification of neuronal network models, also used as the high-level interface to the BrainScaleS hardware}}
\newglossaryentry{MappingTool}{name={MappingTool}, description={The first software framework for mapping neural networks to the BrainScaleS hardware}}
\newglossaryentry{pyhmf}{name={pyhmf}, description={The \gls{PyNN} interface the the BrainScaleS hardware}} %
\newglossaryentry{Marocco}{name={Marocco}, description={The current software framework for mapping neural networks to the BrainScaleS hardware}} %
\newglossaryentry{Spikey}{name={Spikey}, description={Single-chip mixed-signal neuromorphic hardware system - predecessor of the HICANN chip at Electronic Vision(s) group at Heidelberg University}}

\title{Synapse Loss Estimation for the BrainScaleS Wafer-scale Neuromorphic System}
\author{Bernhard Vogginger}
\author{Christian Mayr}

\affil{Chair of Highly-Parallel VLSI-Systems and Neuro-Microelectronics, TUD Dresden University of Technology, Dresden, Germany}
\affil{\textbf{Contact:} bernhard.vogginger@tu-dresden.de}

\begin{document}
\maketitle

\begin{abstract}

Neuromorphic hardware combines memory (synapses) and computation (neurons) into
the same silicon substrate to avoid the von-Neumann bottleneck.
This way, neuromorphic computing aims to make computational neuroscience
simulations and AI processing faster and more energy-efficient.
The common \emph{crossbar} architecture integrates a synapse matrix (analog,
mixed-signal or digital) with a neuron array into a neurosynaptic core, whereas 
multiple cores are interconnected via a dedicated spike routing network.
One of such neuromorphic architecture is the BrainScaleS wafer-scale system
which allows the realization of very flexible network architecture, supporting
both dense and sparse connectivity by highly configurable neurosynaptic cores.
Yet, when network models from computational neuroscience are mapped to the
BrainScaleS wafer, synapse loss can occur which means that for some model
synapses no hardware synapse is available due to the restricted connectivity of
the hardware.
In this work we analyze the synapse loss when mapping uniform random networks
to BrainScaleS both theoretically and empirically.
We first develop a methodology to estimate maximum-sized networks without
synapse loss based on probability distributions and the hardware's synapse
routing architecture.
Next, we adopt the methodology to predict the synapse loss for denser or larger network models.
Then, we compare the predictions with results from actual runs of the mapping software:
The results show the capabilities and limitations of the BrainScaleS system itself and spot
shortcomings of the current mapping algorithms.
The developed methodology can be helpful in multiple ways:
to find the optimal settings when mapping a given network model to BrainScaleS,
to act as a benchmark for the mapping software,
and as tool for design space exploration for new hardware.
While being developed for uniform random networks on the BrainScaleS wafer, our
methodology can be extended to other neuromorphic or in-memory computing
systems and other connection probability distributions, thus improving
model/hardware/software codesign in neuromorphic computing.

\end{abstract}

\section{Introduction}
\label{sec:intro}
By implementing neuron and synapses in dedicated circuits or processors, neuromorphic computing aims to make spiking neural network simulations faster and more energy-efficient.
Besides efficient AI inference and neuro-robotics, scalable brain simulation is one of the main application areas of neuromorphic hardware.
The recent review by \cite{senk2026constructive} shows how different hardware approaches (CPU, GPU, FPGA, digital and mixed-signal neuromorphic) compete to simulate a full-density cortical microcircuit model \cite{potjans2014cell} faster than biological real-time and with least energy.

One promising approach are accelerated, mixed-signal neuromorphic systems that emulate neuron and synapse dynamics in dedicated analog/mixed-signal circuits such as the BrainScaleS systems \cite{schmitt2017neuromorphic,pehle2022brainscales2}.
The BrainScaleS wafer-scale system combines 384 neurosynaptic cores on a silicon wafer and interconnects them with a spike routing grid spanning across the wafer.
To achieve a high synapse density, BrainScaleS employs the common \emph{crossbar architecture} in its neurosynaptic core, where a synapse matrix receives spike input via synapse drivers on the left and right of the matrix and then forwards synaptic events (post-synaptic currents) to an array of neuron circuits below.
To not only support dense networks but also sparse and structured networks, BrainScales uses a very sophisticated spike routing architecture (including address decoders in the synapse) that allows a flexible assignment of model synapses to hardware synapses.
Furthermore, by combining multiple neuron circuits, larger \emph{compound neurons} can be formed with a higher number of input synapses, thus supporting more than \num{10000} synapses per neuron.
Overall, a BrainScaleS wafer contains nearly 200k neurons and about 44~M synapses running \num{10000} faster than biological real-time.

Due to the high complexity of the BrainScaleS routing architecture, a multi-step mapping process is required to map biological network models to the hardware.
During the mapping, it may happen that for a model synapse no suitable hardware synapse can be found, which we call \emph{synapse loss after mapping} or just \emph{synapse loss} \cite{petrovici2014characterization}.
The reason for the synapse loss may either be a structural limitation (e.g., the maximum number of synapses per neuron) or a shortcoming of the mapping software and algorithms.

In this work we analyze the \emph{mapping of uniform random networks to the BrainScales wafer} both analytically and empirically.
We first develop a method to estimate the maximum size of uniform random networks that can be realized on the BrainScaleS wafer without synapse loss.
Then, we extend our method to estimate the synapse loss for larger or denser uniform random networks.
Finally, we compare our predictions to results of actual runs of the mapping software.

This article is structured as follows:
In \cref{sec:methods}, after introducing the BrainScaleS wafer-scale neuromorphic system and the mapping process, we derive the analytical methods to estimate the maximum size of lossless networks and to estimate the synapse loss for denser networks.
\Cref{sec:results} shows the estimations using the developed methods and compares them to empirical results running the actual mapping software.
The results are discussed and interpreted.
Finally, in \cref{sec:discussion} we summarize the results, analyze the chosen design parameters of the BrainScaleS routing architecture, and discuss the potential and limitations of our work in the broader context.

\section{Material and Methods}
\label{sec:methods}

\subsection{BrainScaleS}

\begin{figure}
    \centering
    \includegraphics[width=0.8\linewidth]{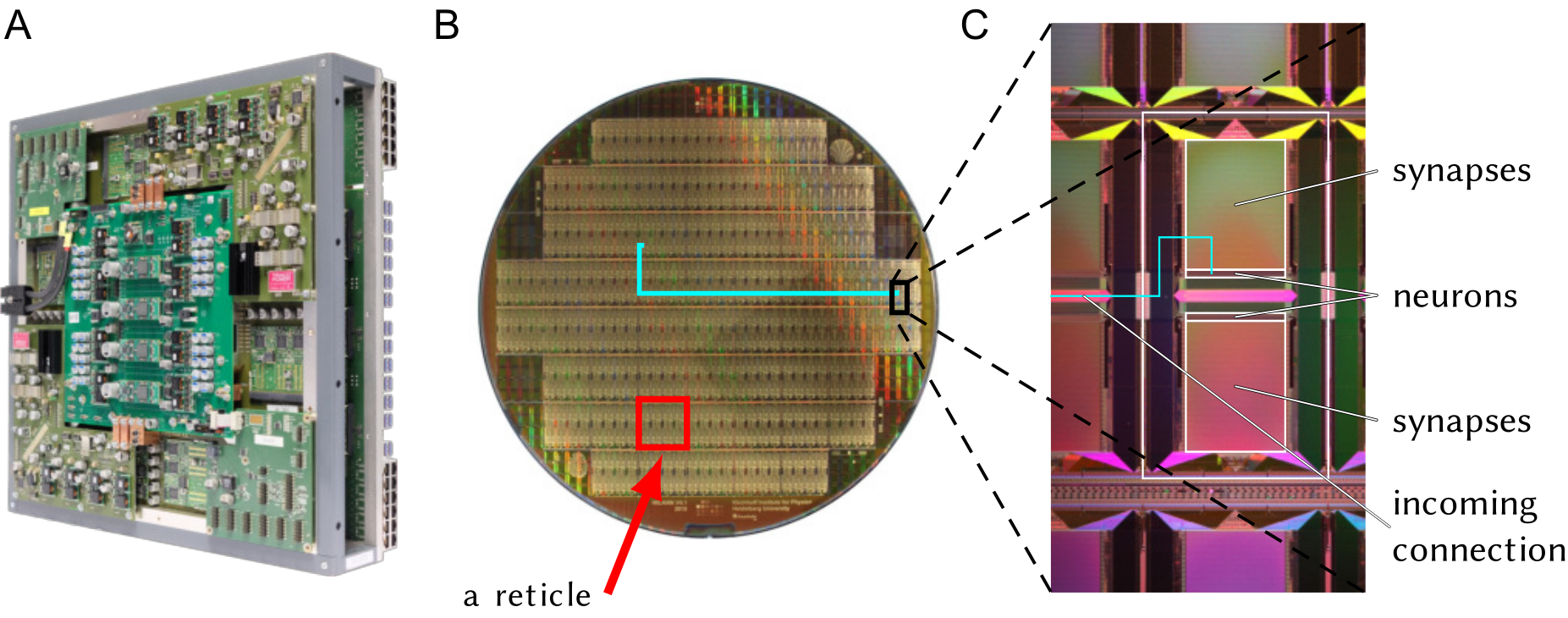}
    \caption{\textbf{BrainScaleS wafer-scale hardware:}
     \textbf{A} Assembled wafer module.
     \textbf{B} Wafer after post-processing with reticles containing 8 HICANNs each.
     \textbf{C} HICANN ASIC with neuron and synapse blocks highlighted. The on-wafer routing network is located left and right of the synapse arrays and between them.
     Figure from \cite[Fig.~1]{kungl2019accelerated} with permission.
    \label{fig:bss-overview}
    }
\end{figure}

In this section we describe the BrainScaleS wafer-scale system with focus on the on-wafer routing network and the related network mapping methodology.

\subsubsection{Hardware description} %
\label{sec:hardware}
\paragraph{System overview}
The BrainScaleS hardware (\Cref{fig:bss-overview}) is built around a \SI{20}{\cm} silicon wafer in UMC 180nm technology containing 384  \glspl{HICANN} \glspl{ASIC}, with each 8 HICANNs per reticle.
After fabrication a post-processing is applied to connect the HICANNs across reticle boundaries and to add electrical contacts for off-wafer connections.
The wafer is mounted onto a printed circuit using elastomer connectors and integrated into to frame (\Cref{fig:bss-overview}A) which contains several other custom boards for the FPGA-based off-wafer communication, for power supply and for readout of analog signals \cite{schmidt2023clean}. 

\begin{figure}
    \centering
    \includegraphics[width=\linewidth]{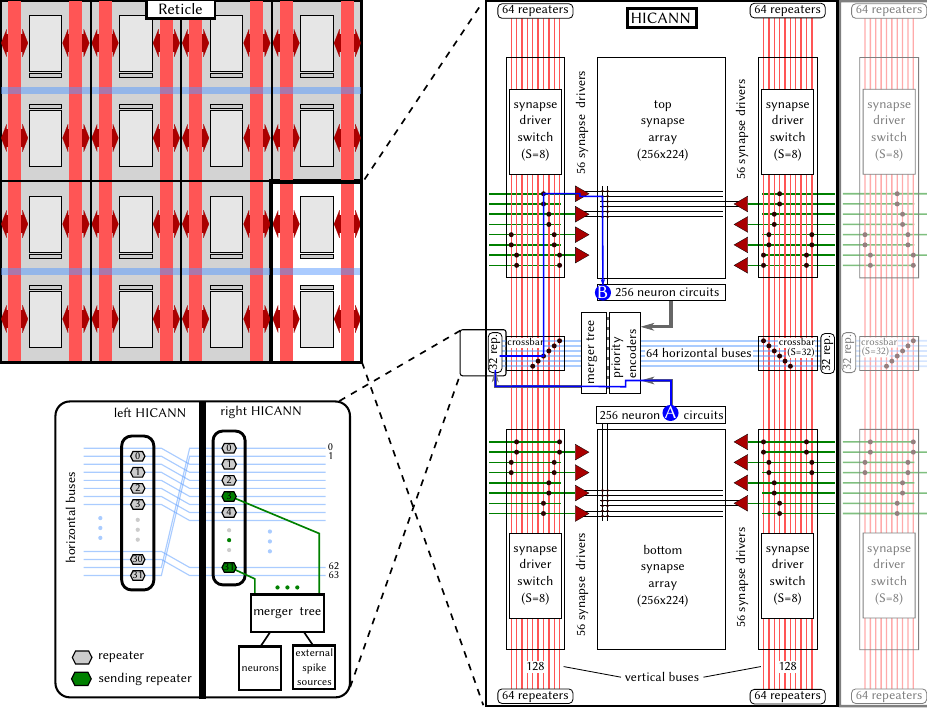}
    \caption{\textbf{On-wafer routing network:} A reticle (top left) contains $4\times2$ HICANNs with a grid of horizontal and vertical buses for spike event routing, which can be connected through sparse crossbar switches. Spikes are detected in neuron circuits, processed by a priority encoder and a merger tree, and then injected with a 6-bit address into a horizontal bus via a sending repeater. The analog serial spike events are re-generated at HICANN edges in repeater circuits. At target HICANNs, the events are routed via further static sparse switches to synapse drivers and thus the synapse arrays. The dark blue line shows an example route from neuron A to neuron B.
    }
    \label{fig:wafer-routing}
\end{figure}

\paragraph{HICANN overview}
Each \gls{HICANN} contains 512 neurons \cite{millner2010} and \num{114688} synapses realized as mixed-signal circuits that emulate the neural dynamics \num{10000} times faster than in biology.
The neuron circuits implement the \gls{AdEx} neuron model \cite{brette05adaptive} with conductance-based synapses and can be configured with individual parameters stored in analog floating-gate memory.
The HICANN is split into an upper and lower block each containing a row of 256 neurons and a synapse array of size $256 \times 224$ (\Cref{fig:wafer-routing}), hence there are 224 afferent (incoming) synapses per neuron.\footnote{Note that within in this paper we consider the HICANNv2 layout with 224 synapses rows per block, while newer wafers have layout HICANNv4 with only 220 synapse rows as in \cite{schmitt2017neuromorphic}.}
Up to 64 neurons (each 32 from top and bottom block) can be interconnected to form larger \emph{compound neurons}, this way supporting up to \num{14336} incoming synapses.

\paragraph{On-wafer spike routing}
The routing of spikes from source neurons to target synapses is realized through a grid of horizontal and vertical buses spanning across the wafer.
Each HICANN has 64 horizontal buses in the center and 128 vertical buses on each side of the synapse array (\Cref{fig:wafer-routing}).
Sparse \emph{crossbar switches} allow to connect horizontal and vertical buses, thus creating fixed multi-cast routes across the wafer.
Connections to the synapse array can be established by sparse switches connecting the vertical buses and the \emph{synapse drivers} (the inputs to the synapse array).
These \emph{synapse driver switches} can also connect to synapse drivers on the neighbor HICANN.
Neighboring synapse drivers can be concatenated to allow multiple synapse drivers to connect to the same vertical bus.

The routing network transmits 6-bit serial pulses representing the spikes of up to 64 source neurons in an \gls{AER}.
When crossing HICANN borders, the 6-bit pulses are re-generated in so-called \emph{repeater} circuits to avoid signal degradation for long routes over the \gls{LVDS} buses.
Initially, the pulses are generated by special \emph{sending repeaters} that receive spike events from the neuron blocks:
After spike detection in the hardware neuron, the 6-bit address assigned to the neuron is first processed by \emph{priority encoders} that handles spikes from 64 hardware neurons, followed by a \emph{merger tree} that routes events from 8 priority encoders to 8 horizontal buses via the sending repeaters.
The merger tree also allows to forward spikes to the off-wafer routing network for recording and to inject external stimulus spikes from there.

\begin{figure}
    \centering
    \includegraphics[width=0.8\linewidth]{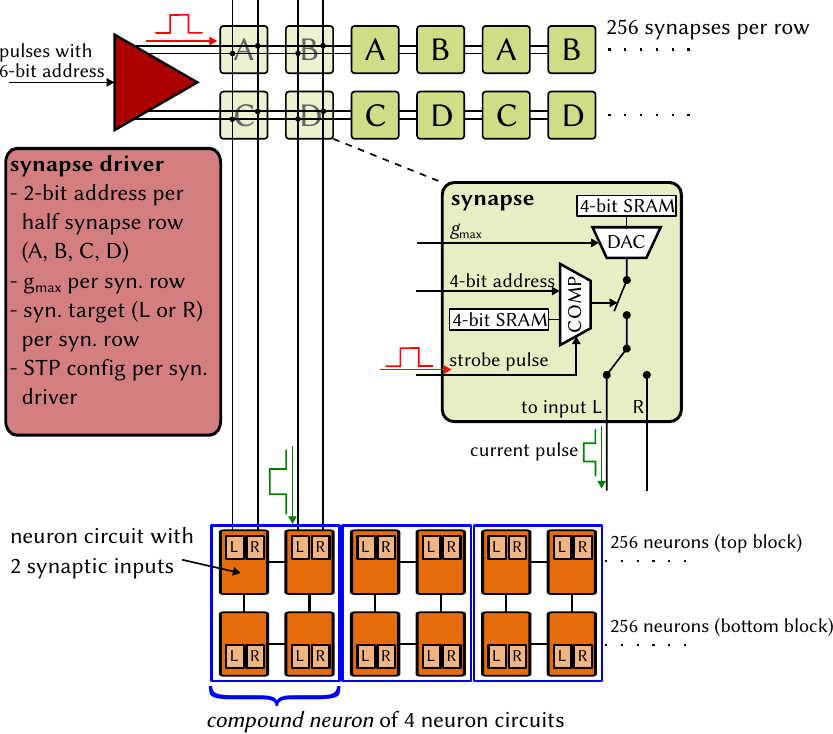}
    \caption{\textbf{Spike processing in synapse array:}
    Spikes arrive as 6-bit pulses in the synapse driver. The \texttt{2 MSB} are compared to the 2-bit configuration per strobe pattern A, B, C, D (also called \emph{half synapse rows}). Upon a match, strobe pulses are sent to the synapse array.
    The pulse height is controlled by parameter $\gmax$ (configurable per synapse row), while the pulse length is determined by a \gls{STP} circuit in the synapse driver.
    In the synapse circuits, the \texttt{4 LSB} of the 6-bit event are compared to a 4-bit value in SRAM.
    In case of a match, a current pulse is generated with its height depending on a 4-bit weight and $\gmax$.
    This current pulse is transmitted to either the left or right synaptic input circuit of the neuron in the same column (configured per synapse row).
    }
    \label{fig:synapse_operation}
\end{figure}

\paragraph{HICANN synapse array}
The processing of incoming spikes events inside the synapse array is described in \cref{fig:synapse_operation}:
The 6-bit addresses representing the source neuron are decoded in two stages: the 2~\gls{MSB} are decoded in the synapse driver and the 4~\gls{LSB} in the synapse circuits.
Each synapse driver is connected to two synapse rows via four strobe lines according to the patterns A, B, C, D in the figure: strobe line A connects to even synapse columns in upper synapse row, B to odd synapses in upper row, etc.
If the \texttt{2 MSB} of the 6-bit address matches the 2-bit address configured per strobe pattern, strobe pulses are generated and sent to the synapse array with amplitude defined by a row-wise scaling factor $\gmax$ and pulse length given by a \gls{STP} circuit in the synapse driver.
In the synapse, the \texttt{4 LSB} of the 6-bit address are compared to a 4-bit decoder address per synapse.
If it matches, the synapse circuit generates a \emph{current pulse} with amplitude depending on a 4-bit weight via a \gls{DAC} and row-wise scaling factor $\gmax$ and duration given by the strobe pulse.

Depending on a row-wise configuration, the current pulse is transmitted to the left or right \emph{synaptic input circuit} of the neuron in the same column.
Each synaptic input circuit implements a conductance-based synapse with configurable reversal potential and decay time constant.
A current pulse from a synapse increases the modeled conductance proportional to the integral of the current pulse.
Typically, the two synaptic inputs are used to model excitatory and inhibitory synapses, but, by combining multiple neuron circuits, it is possible to implement neurons with more than two synapse types, for example to support multiple different excitatory synaptic time constants.

\subsubsection{Mapping process}
\label{sec:mapping}

Here we briefly describe the process of mapping spiking neural networks to the BrainScaleS wafer (see \cite{jeltsch2014phdthesis} for detailed description and \cite{muller2022operating} for the overall software stack):

\paragraph{Network description}
Neural networks are defined in a Python program using the \gls{PyNN} API \cite{davison2009pynn}) by means of \emph{populations} and \emph{projections}.
A \texttt{pyNN.Population} is a group of neurons with the same neuron model but individual parameters. Synapses between two populations are created by a \texttt{pyNN.Projection} offering different connectors (e.g., one-to-one, all-to-all, fixed-probability, distance-dependent, from-list). Synaptic weights can be static or dynamic, then implementing a specific short-term or long-term plasticity rule such as \gls{STDP}.
The synapse model is part of the neuron model, for example, the \texttt{pyNN.if\_cond\_exp} model describes a leaky integrate-and-fire model with exponentially-decaying conductance-based synapses with different parameters for excitatory and inhibitory synapses.
The synapse type (excitatory or inhibitory) is set per projection with the \texttt{receptor\_type} keyword.
PyNN also allows to define external stimuli by connecting Populations of spike source neurons (e.g., Poisson-distributed or used-defined spike times), and the variables to be recorded during simulation, such as spikes or the membrane voltages.

\paragraph{Mapping steps}
The input to the mapping software is the PyNN network description which is converted to a graph of populations with projections represented as edges between them.
Additionally, hardware-specific information can be provided such as the HICANNs to be used, calibration or defect information, as well as  algorithm-specific parameters.

\begin{figure}
    \centering
    \includegraphics[width=0.8\linewidth]{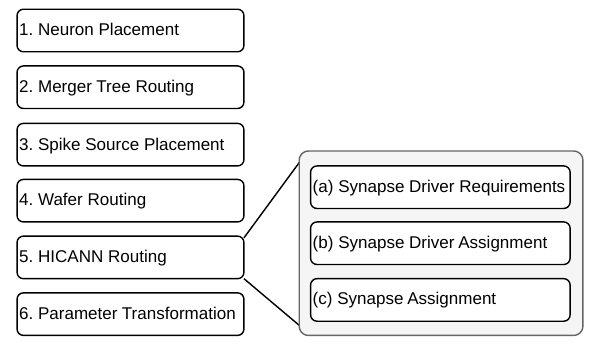}
    \caption{\textbf{Mapping steps}}
    \label{fig:mapping}
\end{figure}

Then, the following mapping steps are performed to generate a configuration for the BrainScaleS hardware:
\begin{enumerate}
    \item \textbf{Neuron Placement:}
    Model neurons are placed to compound neurons on the HICANNs. The size of the compound neurons can be set by the user.
    The placement can be done automatically (e.g., population-wise putting populations onto different HICANNs) or manually in the Python script.
    
    \item \textbf{Merger Tree Routing:}
    Here, the connections between the compound neurons and the on-wafer routing buses are established. 6-bit addresses are assigned to each mapped neuron and the merger tree is configured such that no more than 64 neurons transmit spikes to a horizontal bus. For example, if 128 neurons of size 4 are mapped to a HICANN, 2 horizontal buses will be used.
    
    \item \textbf{External Input Placement:}
    External neurons (spike sources) are assigned to input channels on the wafer (8 per HICANN) with a 6-bit address.
    These input channels connect to horizontal buses via the 8 sending repeaters per HICANN.
    Horizontal buses already connected by real neurons cannot be used anymore. The algorithm tries to minimize the number of horizontal buses used and may consider spike rates to not exceed off-wafer bandwidth limits.
    
    \item \textbf{Wafer Routing:}
    Starting point is the list of horizontal buses per HICANN to which either real neurons or external neurons are mapped to (up to 64 each).
    These \emph{on-wafer routing sources} need to be connected to all respective target HICANNs.
    First, the list of target HICANNs for each routing source is extracted from the network graph and the neuron placement.
    Next, the routes to the target HICANNs are allocated by configuring the repeaters and crossbar switches for connecting the horizontal and vertical buses:
    For this, the routes are first extended horizontally and then vertically across the wafer to reach all target HICANNs.
    It may happen that not all target HICANNs are reached leading to synapse loss.
    
    \item \textbf{HICANN Synapse Routing:}
    Vertical buses are connected to synapse drivers and model synapses are mapped to hardware synapses. This is implemented in three steps:
    \begin{enumerate}
        \item For each vertical bus with target neurons on the HICANN calculate the number of required synapse drivers to realize all model synapses.
        \item Assign and connect synapse drivers to vertical buses via the synapse driver switches and through synapse driver concatenation.
        \item Assign model synapses to hardware synapses by  configuring the respective address decoders in the drivers and synapses.
    \end{enumerate}
    Synapse loss may occur in steps (b) and (c).

    This mapping step is the one most relevant for the studies in this article. Sub-step (a) is described in \cref{sec:synapse_driver_requirements}, for sub-steps (b) and (c) please refer to \cite{jeltsch2014phdthesis}.
    
    \item \textbf{Parameter Transformation:}
    Neuron and synapse model parameters are translated to a configuration for the hardware circuits.
    Model parameters are first translated to the hardware domain considering the time and voltage scales of the circuits.
    Then calibration data is applied to obtain the 10-bit DAC vales for the neuron parameters.
    Finally, 4-bit synapse weights and row-wise $\gmax$ settings are calculated based on calibration data and optimization mechanisms to minimize quantization and clipping errors.    
\end{enumerate}
An overview of the mapping steps is shown in \cref{fig:mapping}.

\subsection{Synapse driver requirements}
\label{sec:synapse_driver_requirements}
In this subsection we introduce the \emph{synapse driver requirements calculation}, the first part of the HICANN Routing step. This is necessary to understand the newly developed methods in later sections: maximum network size estimation in \cref{sec:method-max-sources} and synapse loss estimation for uniform random networks networks in \cref{sec:expected_loss}. We first introduce terms and prerequisites in \cref{sec:prerequesites}, then describe the algorithm used in the software stack in \cref{sec:syndriver_calculation}, and close with deriving the number of required synapse drivers for one routing source for uniform random connectivity (\Cref{sec:method-max-sources}).

\subsubsection{Prerequisites}
\label{sec:prerequesites}
On the highest level, synapse drivers with its 2 synapse rows of total 512 synapses are the limited resources for the HICANN synapse routing because it is the smallest entity to be connected to one vertical bus.
The \emph{synapse driver requirements} are calculated for each HICANN side, considering the on-wafer routing sources available on the vertical buses either left or right of the HICANN.
We assume a neuron size of at least 2, such that neurons from upper and lower block are always connected and such that each side can be treated as a whole \cite{fieres2008realizing}.

For actual network topologies, the number of required synapse drivers per HICANN side is calculated as follows (simplified, neglecting short-term plasticity):
Find all on-wafer routes available on the vertical buses next to the HICANN side carrying signals from neurons that have model synapses to neurons mapped to this HICANN.

\paragraph{Model synapses}
For each on-wafer route extract the model synapses to neurons on this HICANN.
The \emph{model synapses} are defined by:
\begin{itemize}
    \item \Addr: 6-bit address of source neuron
    \item \TgtNrn: target neuron (Index of compound neuron on HICANN)
    \item \SynType: synapse type
    \item synaptic weight
    \item \gls{STP} parameters
\end{itemize}
For simplicity we ignore the weight for now and assume static synapses without STP so that model synapses are definded by:

\begin{equation}
S_\text{model} = S_\text{model} (\Addr, \TgtNrn, \SynType)
\end{equation}

\paragraph{Hardware synapses}
Instead, \emph{hardware synapses} (cf.~\cref{fig:synapse_operation}) are defined by
\begin{itemize}
    \item 4-bit decoder address of the synapse
    \item 2-bit decoder address assigned to the \emph{half synapse row}
    \item Whether the synapses of the \emph{synapse row} connect to the \emph{left} or \emph{right} synaptic input circuit of the neuron
    \item $\gmax$ scaling factor per \emph{synapse row}
    \item \gls{STP} hardware parameters
\end{itemize}
Again, let's ignore the weight scale and STP settings for now.

\paragraph{Mapping problem}
Hence, in order to map model synapses to hardware synapses, the following is needed:
\begin{enumerate}
    \item The synaptic input circuit of the connected hardware neuron must implement the synapse type of the synapse.
    \item 2-bit decoder address of half synapse row pattern matches the 2-MSB of the source address
    \item 4-bit synapse decoder address matches the 4-LSB of the source address
\end{enumerate}

Obviously, model synapses with the same target neuron, synapse type and 2-MSB address are equivalent in the mapping process, that is, they can be implemented by same set of hardware synapses.

\paragraph{Synapse target mapping}
Before calculating the number of required synapse drivers, we first perform a mapping of synapse types to input circuits of the hardware neurons.
Here, we assume the \emph{simple case} with just \emph{two synapse types} and map \emph{excitatory to the left} inputs and \emph{inhibitory to the right} inputs in each neuron column.
As a consequence, all synapses in one row can implement either excitatory or inhibitory synapses.

\paragraph{Synapses per half synapse row}
As shown in \cref{fig:synapse_operation}, compound neurons can be formed with up to 64 hardware neurons.
If the neuron width is larger than 2, there are multiple synapses belonging to the same strobe pattern that connect to same compound neuron.
\Cref{tab:syns_per_half_row} lists the number of synapses per pattern (here called \emph{half synapse row}) depending on the neuron size.
For example, in case of neuron size 8, when assigning one half synapse row to MSB pattern \texttt{0b00}, 2 synapses from that address range \texttt{[0,15]} can be implemented.

\begin{table}[h]
    \centering
    \begin{tabular}{l|cccccc}
      Neuron size   & 4 & 8 & 12 & 16 & 32 & 64 \\
      \hline
      Neuron width  & 2 & 4 & 6 & 8 & 16 & 32 \\
      Synapses per half row & 1 & 2 & 3 & 4 & 8 & 16
    \end{tabular}
    \caption{\textbf{Number of synapses per half synapse row}}
    \label{tab:syns_per_half_row}
\end{table}

\subsubsection{Synapse driver requirements calculation}
\label{sec:syndriver_calculation}
The number of required synapse drivers per routing source are calculated as follows:

\begin{enumerate}
    \item \textbf{Synapses per pattern and type:} For each target neuron $i$, count the number model synapses per 2-MSB pattern and synapse type, yielding:
    \begin{align}
	   S^i(\Decoder,\SynType) \quad .
    \end{align}
    Note, practically, $S^i(\Decoder,\SynType)$ is a 3D array of shape $(\Nnrn, 4, 2)$ with the synapse counts for all $\Nnrn$ target neurons, the four 2-MSB patterns, and two synapse types. 
    
    \item \textbf{Half rows per pattern and type:} For each target neuron $i$, calculate the number of half synapse rows required to realize all synapses per 2-MSB pattern and synapse type considering the neuron size and \Cref{tab:syns_per_half_row}.
    \begin{align}
	   \HalfRows^i(\Decoder,\SynType) = \left \lceil \frac{S^i(\Decoder,\SynType)}{\textrm{synapses per half row}} \right \rceil \quad .
    \end{align}
    
    \item \textbf{Half rows per pattern and type (all neurons):} Next, we take the maximum number of half rows across all neurons for each 2-MSB pattern and synapse type.
    \begin{align}
    	\HalfRows(\Decoder,\SynType) &= \max\limits_{i}(\HalfRows^i(\Decoder,\SynType)) \label{eq:half-row-nrn}
    \end{align}
    
    \item \textbf{Synapse rows per side:} Next, we consider the synapse type and calculate the number of synapse rows $\Rows(\Side)$ required per input target (left or right). For this, for each synapse type we compute the sum of the half rows over the patterns, divide it by 2 and round up:
    \begin{align}
	\Rows(\Side\!=\!\textrm{left}) &= \left \lceil \frac{1}{2} \sum_\textrm{patterns} \HalfRows(\Decoder, \SynType\!=\!\textrm{excitatory}) \right \rceil \quad, \label{eq:rowsperside}\\
    	\Rows(\Side\!=\!\textrm{right}) &= \left \lceil \frac{1}{2} \sum_\textrm{patterns} \HalfRows(\Decoder, \SynType\!=\!\textrm{inhibitory}) \right \rceil \nonumber
    \end{align}

    \item \textbf{Synapse drivers:} Finally, calculate the number of drivers by summing up the rows per side, dividing by 2 and rounding to the ceiling:
    \begin{align}
        \Drivers =  \left \lceil \frac{1}{2} \sum_\textrm{sides} \Rows(\Side) \right \rceil \quad .
    \end{align}
\end{enumerate}
Based on this, the total number of drivers per HICANN side is calculated by summing the requested drivers of all on-wafer routing sources.

Note that the described approach is a simplification of the actual implemented software algorithm for synapse driver requirements calculations which also considers the \gls{STP} mode (off, facilitating, depression). It also supports neuron models with more than 2 synaptic input circuits and flexible synaptic target mappings.

\subsubsection{Synapse driver estimation for uniform random networks}
\label{sec:method-syndriver-estimation}
The required number of routing resources for uniform random networks can be estimated with the following, newly derived methodology.

\paragraph{Binomial distribution applied to uniform random networks}
When connecting $n$ source neurons with connection probability $p$ to a single target neuron, the number of synapses follows the binomial distribution:
\begin{equation}
P(X = k) = B(k|n, p) = \binom{n}{i} p^k (1-p)^{n-i} \quad , \label{eq:binom}
\end{equation}
(The probability that there are $k$ synapses equals $B(k|n, p)$).

Accordingly, the cumulative distribution of the binomial distribution
\begin{equation}
	F(k|n,p) = P(X \leq k) = \sum_{i=0}^{\lfloor k \rfloor} \binom{n}{i} p^k (1-p)^{n-i} \label{eq:binomcdf}
\end{equation}
can be used to calculate the probability that there are less or equal $k$ synapses.

If we consider a uniform random network with $n$ source neurons and $\Nnrn$ target neurons with connection probability $p$, we can calculate the probability that for all target neurons $\Nnrn$ there are less or equal than $k$ synapses:
\begin{equation}
	F(k|n,p,\Nnrn) = P(X \leq k \text{ for all } \Nnrn) = F(k|n,p)^{\Nnrn} \label{eq:cdfallneurons} \quad.
\end{equation}
Explanation: The probabilities for all target neurons are independent and identically distributed, hence the probabilities need to be multiplied.

\paragraph{Required synapse drivers per routing source}
We extend this approach to the estimate the number of required hardware resources per routing source. For this, we assume a homogeneous synapse target mapping, hence all neuron sizes on one HICANN are the same. Furthermore, we assume that source neurons on one routing source are either all excitatory or inhibitory, such that we don't need to distinguish between synapse types when calculating the number of synapse rows as in \cref{eq:rowsperside}.

\Cref{eq:cdfallneurons} provides the formula that all target neurons have less or equal than $k$ synapses.
We make use of this to calculate the probability that the maximum number of synapses $S_\mathrm{max}$ to all $\Nnrn$ target neurons is $k$:
\begin{equation}
	P_{S_\mathrm{max}}(k) = P(\max_{i=1..\Nnrn} S_i\!=\!k) = 
	\begin{dcases}
		F(0|n,p,\Nnrn),                     &\text{for } k=0 \\
		F(k|n,p,\Nnrn) -  F(k-1|n,p,\Nnrn), &\text{if } k\geq 1.
	\end{dcases}
	\label{eq:probmaxk}
\end{equation}
Explanation: The probability for $S_\mathrm{max}\!=\!0$ is given by $F(0|n,p,\Nnrn)$; for higher $k$ values the probability can be calculated using the difference of $F(k|n,p,\Nnrn)$ and $F(k-1|n,p,\Nnrn)$.

Next, we look at synapses from one 2~MSB pattern (hence, 16 source neurons) and consider the number of synapses per half synapse row $\SynsPerHalfRow$ (which depends on the neuron size, see \cref{tab:syns_per_half_row}) to calculate the probability that $k$ half synapse rows are required to realize all synapses:
\begin{equation}
	P_{H,\mathrm{pattern}}(k) =
	\begin{dcases}
		P_{S_\textrm{max}}(0),                                                          &\text{for } k=0\\
		\sum_{i=\SynsPerHalfRow(k-1) + 1}^{\SynsPerHalfRow k+1} P_{S_\textrm{max}}(i) , &\text{for } k\geq 1
	\end{dcases}
	\label{eq:prob_halfrows_pattern}
\end{equation}
Note that above formula provides probabilities for \cref{eq:half-row-nrn} in the requirements calculation.

Next, we combine the probabilities of all four 2~MSB patterns to calculate the probability that $k$ half synapse rows are required to fit all synapses from the 64 source neurons of one on-wafer routing source:
\begin{equation}
	P_{\mathrm{H}}(k) = \sum_{ \{k = k_1+k_2+k_3+k_4\}} \prod_{i=1}^4 P_{H,\mathrm{pattern}_i}(k_i),
	\label{eq:prob_halfrows_l1source}
\end{equation}
where $\{k = k_1+k_2+k_3+k_4\}$ is the set of all tuples $(k_1,k_2,k_3,k_4)$
whose elements sum up to $k$.
For example, for the probability for $k=1$ half rows one has to consider 4 cases, namely the tuples $(1,0,0,0)$, $(0,1,0,0)$, $(0,0,1,0)$, $(0,0,0,1)$.\footnote{Note that the software implementation used for the results in \cref{sec:results} considering the HICANN constraints (\Cref{sec:hardware-limitations}) supports that the probabilities per pattern $P_{H,\mathrm{pattern}_i}$ can be different.}

Finally, we calculate the probability that $k$ synapses drivers are required for realizing all synapses by taking into account that there are 4 half synapse rows per driver:
\begin{equation}
	P_D(k) =
	\begin{dcases}
		P_H(0),                             &\text{for } k=0\\
		\sum_{i=4(k-1) + 1}^{4k+1} P_H(i) , &\text{for } k\geq 1 .
	\end{dcases}
	\label{eq:probdrivers}
\end{equation}
Explanation: The probability for requiring 1 driver is equal to the sum of the probabilities for requiring 1 to 4 half synapse rows, etc.

\begin{figure}
	\includegraphics{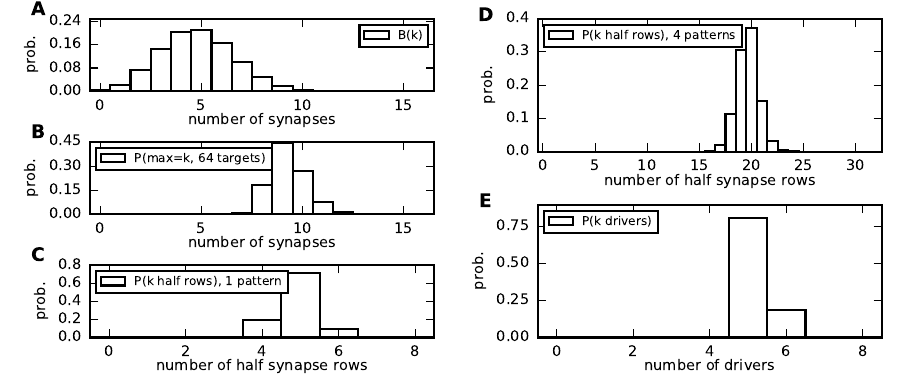}
	\caption{
		\textbf{Example: Required resources for uniform random networks.}
    \small
		This figure illustrates how to compute the number of required hardware
		resources to realize all synapses from an on-wafer routing source to
		all target neurons on a HICANN.
		Parameters used here: connection probability $p=0.3$, $\Nnrn=64$ target
		neurons per HICANN of size 8.
		\textbf{A} The number of input synapses per target follows a binomial
		distribution ($n=16$ sources per 2 MSB address pattern).
		\textbf{B} Probability that $k$ is the maximum number of synapses per
		target for all $\Nnrn$ target neurons.
		\textbf{C} Probability that $k$ half synapse rows are required to
		realize all synapses from the $n=16$ sources per 2 MSB address pattern.
		\textbf{D} Number of half synapse rows required for all four 2 MSB patterns.
		\textbf{E} Number of required synapse drivers.
		\label{fig:expected_half_rows}
	}
\end{figure}

\Cref{fig:expected_half_rows} show an example for estimating the required resources for one on-wafer routing source following the approach described above.

\subsection{Estimating the maximum number of source neurons per HICANN}
\label{sec:method-max-sources}
Based on \cref{eq:probdrivers} one can calculate the \emph{expected number of drivers} required to realize all synapses per on-wafer routing source:
\begin{equation}
	\mathbb{E} [D] = \sum_k k P_D(k) \quad . \label{eq:expected-drivers}
\end{equation}
Note that this is an expectation value, not a discrete number.

Next, we calculate how many routing sources can be served with the 224 synapse drivers per HICANN considering the expected number of drivers per source (\Cref{eq:expected-drivers}) required to implement all synapses.
The \emph{expected number of routing sources} is given by:
\begin{align}
	\mathbb{E} [\text{routing sources}] &= \min \left (\frac{224}{\mathbb{E}[D]}, 224 \right) \quad , \label{eq:expected_l1sources}
\end{align}
considering that there cannot be more routing sources than drivers.

Last, the maximum number of input neurons per HICANN for a given connection probability can be estimated by multiplying the number from \cref{eq:expected_l1sources} with the 64 addresses per routing source:
\begin{align}
	\mathbb{E} [\text{source neurons}] &= \mathbb{E} [\text{routing sources}]\cdot 64 \label{eq:expected_sources}.
\end{align}
Note that for this calculation we neglect several hardware constraints, such as that the synapse drivers are separated into 4 blocks, or that the synapse driver switches are sparse which might hinder some connection setups.

\subsection{Synapse loss estimation}
\label{sec:expected_loss}

To estimate the synapse loss for given uniform random networks (either feed-forward or recurrent), we take a reverse approach:
We step by step assign resources to requests and then estimate the fraction of synapses that can not be realized.
First, synapse drivers are assigned to routing sources, then half synapse rows are assigned to the 2 MSB patterns of the 64 addresses on one routing source.
To perform this top-down approach in optimal way, we first have to introduce the concept of the \emph{synapse gain per hardware resource}:

\paragraph{Probability for more than $k$ synapses per neuron}
We go bottom-up and analyze how many model synapses per 2~MSB pattern can be realized for a given number of half synapse rows.
For uniform random networks with connection probability $p$, we consider the survival function $G(k|n,p)$ to calculate the probability that there are more than $k$ synapses per neuron from $n$ source neurons:
\begin{align}
	G(k|n,p) = P(X > k) = 1 - F(k|n,p), \label{eq:binomsf}
\end{align}
with $F(k|n,p)$ being the cumulative distribution from \cref{eq:binomcdf}.

\paragraph{Synapse gain per half synapse row}
Next, we calculate how many additional synapses can be implemented with each additional half synapse row assigned to a 2~MSB pattern for $\Nnrn$ target neurons:
The \emph{synapse gain} $\Delta S_\mathrm{pattern}(k)$ of the $k$th assigned half synapse row is calculated as:
\begin{align}
	\Delta S_\mathrm{pattern}(k) = \sum_{i=k\cdot\SynsPerHalfRow}^{(k+1)\cdot\SynsPerHalfRow-1} G(i|n,p)\cdot \Nnrn , \label{eq:synapse_gain}
\end{align}
which depends on the chosen neuron size through the number of synapses per half synapse row $\SynsPerHalfRow$ (cf.~\cref{tab:syns_per_half_row}).
The number of target neurons $\Nnrn$ is just a multiplicative factor.
Note the zero-based numbering of $k$ in \cref{eq:synapse_gain} such that $k=0$ refers to the first assigned half synapse row.

\paragraph{Realized synapses for $D$ assigned drivers}
Next, we estimate the number of synapses that can be realized when $D$ synapse drivers are assigned to a routing source.
For this, the 4 half synapse rows per driver (A, B, C, D in \cref{fig:synapse_operation}) need to be distributed to the 2 MSB patterns of the 64 addresses per routing bus.
We pursue an optimal distribution of half synapse rows to 2~MSB patterns which maximizes the overall number of realized synapses.
For this, we sort the synapse gains $\Delta S_{\mathrm{pattern},i}(k)$ of all patterns $i$ in descending order:
\begin{align}
	\Delta S_\mathrm{sorted} = \text{sort}_{\geq} \left (\{\Delta S_{\mathrm{pattern},i}(k)| 1 \leq i \leq 4, k \geq 0\}\right ). \label{eq:synapse_gain_sorted}
\end{align}
Then, the number of synapses that can be realized with $D$ assigned synapses drivers per routing source is given by:
\begin{align}
	S_\mathrm{realized}(D) &= \sum_{i=0}^{4\cdot D-1} \Delta S_\mathrm{sorted}(i), \label{eq:realized_syns}
\end{align}
which considers that there are 4 half synapse rows per driver.
Correspondingly, the number of lost synapses amounts to:
\begin{align}
	S_\mathrm{lost}(D)     &= \sum_{i=4\cdot D}^{\infty} \Delta S_\mathrm{sorted}(i) .\label{eq:lost_syns}
\end{align}
Based on this one can calculate the expected relative synapse loss $\nu_\mathrm{loss}$ as
\begin{align}
	\nu_\mathrm{loss} = \frac{S_\mathrm{lost}}{S_\mathrm{realized} + S_\mathrm{lost}}. \label{eq:expected_loss}
\end{align}

\paragraph{Distribution of synapse drivers to routing sources}
In the synapse loss estimation in \cref{sec:results_synloss}, where we sweep the number of routing sources per HICANN, we need to distribute the 224 overall drivers to the routing sources.

When the number of routing sources $L \leq 224$, then we assign:
\begin{itemize}
    \item $D_\mathrm{ceil} = \lceil 224/L \rceil$ drivers to $L_\mathrm{ceil} = (224\bmod L)$ routing sources, and 
    \item $D_\mathrm{floor} =\lfloor 224/L \rfloor$ drivers to the remaining $L_\mathrm{floor} = L - L_\mathrm{ceil}$ routing sources.
\end{itemize}

Then, the total number of realized and lost synapses is calculated as:
\begin{align}
	&S_\mathrm{realized,tot}(L) = L_\mathrm{ceil} \cdot S_\mathrm{realized}(D_\mathrm{ceil}) +  L_\mathrm{floor} \cdot S_\mathrm{realized}(D_\mathrm{floor}) \\
	&S_\mathrm{lost,tot}(L)     = L_\mathrm{ceil} \cdot S_\mathrm{lost}(D_\mathrm{ceil}) +  L_\mathrm{floor} \cdot S_\mathrm{lost}(D_\mathrm{floor})
\end{align}
When the number of routing sources $L$ exceeds the number of synapse drivers, then for $L-224$ routing sources all synapses are lost while the others receive one driver:
\begin{align}
	&S_\mathrm{realized,tot}(L) = 224 \cdot S_\mathrm{realized}(1) \\
	&S_\mathrm{lost,tot}(L)     = 224 \cdot S_\mathrm{lost}(1) + (L- 224) \cdot S_\mathrm{lost}(0)
\end{align}

\subsection{Hardware limitations}
\label{sec:hardware-limitations}
The BrainScaleS wafer-scale system has several limitations due to its analog/mixed signal implementation which affect the mapping and routing of spiking neural networks.
Here, we list those limitations relevant for the studies in this article:
\begin{enumerate}
    \item \textbf{4-bit weight to disable synapses}: The hardware synapses use a 4-bit digital weight which is translated into a current amplitude via a \gls{DAC}.
    Apparently, a digital weight \texttt{0} leads to a non-zero positive current, as has been studied in \cite{koke2017phdthesis}.
    Hence, in order to disable a hardware synapse one 4-bit address (e.g. \texttt{15}) is reserved, which then is never used during the mapping process when assigning 6-bit addresses to neurons on routing buses. This decreases the number of neurons per routing source from 64 to 60.
    \item \textbf{6-bit address \texttt{0} for repeater locking}: The on-wafer routing network uses a \gls{DLL} in the repeaters for recovering the timing of the serial events (6-bit address plus start and stop bit) \cite{schemmel_iscas2010}.
    For a reliable operation events with the 6-bit address \texttt{0b000000} must be sent regularly for locking the DLLs, which is realized through the background event generators.
    This makes the 6-bit address \texttt{0} unusable for real neurons and thus further reduces the \emph{number of usable 6-bit addresses} to 59.
    \item \textbf{Synapse driver concatenation}: As described in \cref{sec:hardware}, the inputs of multiple neighboring synapse drivers can be interconnected to receive the same signal from the synapse driver switches. The length of such synapse driver chains is limited: if too many are connected, the signal quality suffers leading to errors when capturing the 6-bit addresses \cite{kononov11diplomathesis}.
    While the mapping software \cite{muller2022operating} limits this chain length to 3 per default, here we consider 5 as the maximum synapse driver chain length in \cref{sec:results}.
    \item \textbf{Usable HICANNs:} From the 384 HICANNs on the wafer there are 16 that in practice cannot be used for real experiments, because the two reticles they belong don't have off-wafer connections for sending and receiving spikes \cite{schmidt2023clean}. This reduces the number of \emph{usable HICANNs} to 368, which we consider for the synapse loss estimation in \Cref{sec:results_synloss}.
\end{enumerate}

\subsection{List of symbols and constants}
\Cref{tab:symbols} shows a list of symbols and constants used in this article.
\begin{table}[]
    \centering
    \begin{tabular}{c|l|c}
         \textbf{symbol} & \textbf{description} & \textbf{value / value range} \\
         \hline %
         $p$ & connection probability & \\
         $P(X)$ & probability distribution of random variable $X$ & \\
         $F(k)$ & cumulative distribution function $P(X <=k)$ & \\
         $G(k)$ & survival function $P(X > k)$ & \\
         \hline %
         $\Synapses$ & synapses & - \\
         $\HalfRows$ & half synapse rows & - \\
         $\Rows$ & synapse rows & - \\
         $\Drivers$ & synapse drivers & - \\
         $\Nnrn$ & target neurons on HICANN & - \\
         $\TgtNrn$ & index of target neuron on HICANN & \\
         $\SynsPerHalfRow$ & synapses per half synapse row per target neuron & see \cref{tab:syns_per_half_row} \\
         $\Addr$ & 6-bit source address & $[0, 63]$\\
         $\Decoder$ & 2~MSB pattern of $\Addr$ & $[0, 3]$ \\
         $\SynType$ & synapse type & $\{\text{excitatory}, \text{inhibitory}\}$ \\
         $\Side$ & side of synaptic input circuit in hardware neuron & $\{\text{left}, \text{right}\}$ \\

         $U_\text{Syn}$ & synapse utilization & \\
        \hline
         $N_\text{Drv,quadrant}$ & synapse drivers per HICANN quadrant & $\num{56}$ \\
         $N_\text{Drv,HICANN}$ & synapse drivers per HICANN & $\num{224}$ \\
         $N_\text{Nrn,HICANN}$ & hardware neurons per HICANN & $\num{512}$ \\
         $N_\text{Syn,HICANN}$ & hardware synapses per HICANN & $\num{14336}$ \\
    \end{tabular}
    \caption{\textbf{List of symbols and constants}}
    \label{tab:symbols}
\end{table}

\section{Results}
\label{sec:results}

\subsection{Maximum size of lossless uniform random networks}
We apply the methods from \cref{sec:method-syndriver-estimation} and \cref{sec:method-max-sources} to estimate the maximum realizable uniform random networks on the BrainScaleS hardware.

\paragraph{Maximum number of source neurons per HICANN}
\begin{figure}
	\includegraphics{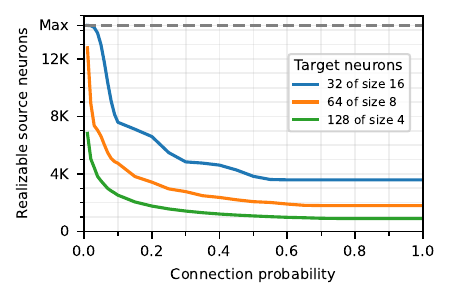}
	\includegraphics{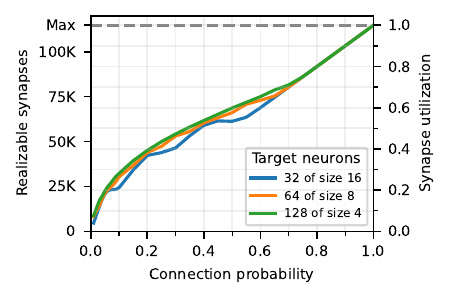}
	\caption{
		\textbf{Realizable uniform random networks on HICANN.}
    \small
		Left: Expected number of source neurons for which all synapses to
		target neurons on a HICANN can be realized depending on the connection
		probability and different mapping scenarios (Number of target neurons
		and neuron size).
		The upper limit of source neurons is given as $224 \text{
			drivers}\times 64\text{ addresses} = \num{14336}$.
		Right: Expected number of realized synapses per HICANN for the data
		from the left according to \cref{eq:expected_synapses}.
		The right axis shows the relative utilization of the overall
		\num{114688} synapses per HICANN.
		\label{fig:sources_vs_pconn}
	}
\end{figure}

We first do this for feed-forward networks on \emph{one HICANN} for three different settings of the compound neuron size (4, 8, or 16 neuron circuits) and thus different number of target neurons (128, 64, 32 respectively).
We sweep the connection probability between $0.01$ and $1.0$ and estimate the number of source according to \cref{eq:expected_l1sources}, as shown in \cref{fig:sources_vs_pconn}.
Two trends can be identified:
First, the larger the neuron size, the larger is the expected number of source neurons supported, which is due to more synapses per half row per target neuron.
Second, the realizable number of source neurons decreases with the connection density, as expected.
Note that in all setups we assume 64 input neurons per on-wafer routing source.
Only for a large neuron size (16) and a connection probability below $0.05$ the number of realizable source neuron reaches the maximum number (number of synapse drivers $\times$ 64).

To estimate the \emph{utilization of hardware synapses}, we first calculate the expected number of synapses as:
\begin{align}
	\mathbb{E} [\text{synapses}] &= \mathbb{E} [\text{source neurons}]\cdot \Nnrn \cdot p\label{eq:expected_synapses},
\end{align}
and then calculate the synapse utilization by dividing by the number of hardware synapses per HICANN:
\begin{align}
    U_\text{Syn} & = \frac{\mathbb{E} [\text{synapses}]}{N_\text{Syn,HICANN}} \quad. \label{eq:syn-util}
\end{align}
The results are shown on the right side of \cref{fig:sources_vs_pconn}:
For $p>0.7$ the synapse utilization is proportional to $p$, reaching \SI{100}{\%} for $p=1.0$.
Instead, for lower connection probabilities, the utilization scales better than linearly.
This scaling behavior is similar for all 3 settings but one can recognize discretization effects for larger neuron sizes: The number of synapses per half row is higher for larger neuron sizes (\Cref{tab:syns_per_half_row}), which leads to more unused hardware synapses for neuron sizes of 16 or 32, and thus a worse overall synapse utilization.

\paragraph{Comparison to hardware architecture with 6-bit address decoding in the synapse} \label{sec:comp_arch}
\begin{figure}
	\includegraphics{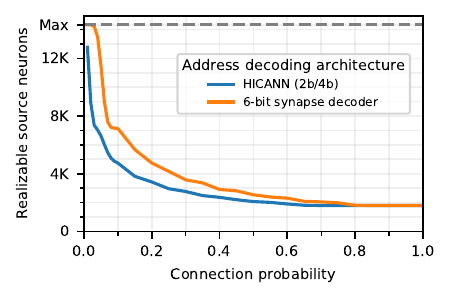}
	\includegraphics{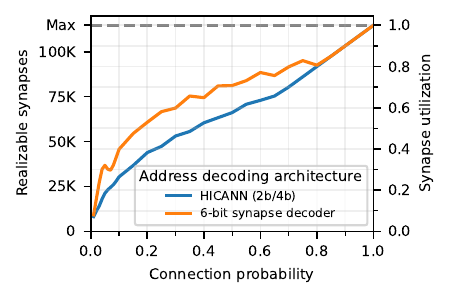}
	\caption{
		\textbf{Comparison to hardware architecture with 6-bit synapse address decoder.}
    \small
		The HICANN decodes the upper 2 bits of the 6-bit neuron addresses for 
		each half synapse row and only 4 bits in the synapse.
		This approach is compared to a setup where all 6 bits are decoded in the synapse.
		Left: Expected number of source neurons for which all synapses to
		target neurons on a HICANN can be realized for 64 target neurons of size 8.
		Right: Expected number of realized synapses per HICANN and synapse utilization.
		\label{fig:sources_vs_pconn_comp_arch}
	}
\end{figure}

The HICANN applies a distributed decoding of the 6-bit source addresses: The 2~MSB are compared per half synapse row in the driver, while the 4~LSB are configurable per hardware synapse.
This is a trade-off between routing flexibility and hardware cost, as each bit in the synapse increases the required silicon area \cite{schemmel_ijcnn2008,fieres2008realizing}.
Here, we analyze the case where all 6 bits are decoded in the synapses.
The remaining conditions stay the same, such as 2 synapse rows per driver.
The expected number of routing sources per HICANN and the synapse utilization are shown in \cref{fig:sources_vs_pconn_comp_arch} for the setup with 64 target neurons of size 8 and compared to the split 2~bit/4~bit decoding:
Especially for lower connection probabilities the number of realizable source neurons is higher for the 6-bit synapse decoder case, which also leads to a significantly higher synapse utilization shown on the right.
Of course, this comes at the price of additional hardware cost (silicon area): 2 more lines for broadcasting the 2~MSB to the synapses as well as 2 additional SRAM cells and 2 extra comparators per synapse are needed.
For example, the height of the synapse circuits would increase, which in turn would reduce the number synapse rows fitting into the same chip area.
Our methodology can be applied to find the optimal hardware architecture for implementing certain networks of desired size and connection density.
We note that the 6-bit address decoding has been applied in the BrainScaleS-2 system \cite{pehle2022brainscales2}, which is implemented in a smaller technology (\SI{65}{\nano\meter} vs.~\SI{180}{\nano\meter}) with reduced area per SRAM cell and digital logic.

\paragraph{Considering all HICANN constraints}

\begin{figure}
	\includegraphics{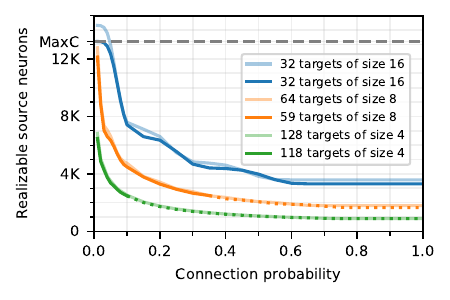}
	\includegraphics{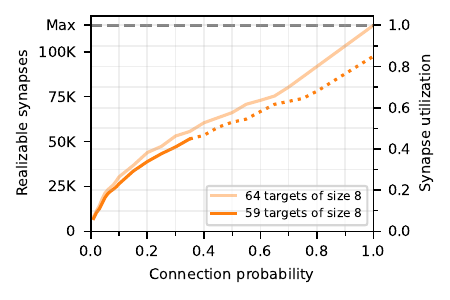}
	\caption{
		\textbf{Realizable uniform random networks on HICANN with constraints.}
    \small
		In real experiments, only 59 out of 64 addresses can be used for real
		source neurons, as described in \cref{sec:hardware-limitations}.
		We compare the realizable uniform networks with 59 sources (darker lines) to the previously assumed 64 sources (lighter lines) per
		on-wafer routing source.
		Left: Expected number of source neurons for which all synapses to
		target neurons on a HICANN can be realized depending on the connection
		probability and different mapping scenarios (Number of target neurons
		and neuron size).
		\texttt{MaxC} denotes the maximum number of source neurons in the
		constrained case: $224 \text{ drivers} \times 59 \text{ addresses} =
		13216$.
		Right: Expected number of realized synapses and synapse utilization per HICANN for neuron size 8.
        The \emph{dotted lines} show the cases affected by the synapse driver chain length limitation: here, all cases are marked where the probability is higher than  \SI{80}{\%} that more than 5 drivers per routing source are required.
		\label{fig:sources_vs_pconn_comp_constraints}
	}
\end{figure}
Next, we analyze the effect of the additional hardware constraints (\Cref{sec:hardware-limitations}) such that only 59 of 64 addresses are usable per on-wafer route and that the number of synapse drivers that can be connected in a chain is limited.
The results are shown in \cref{fig:sources_vs_pconn_comp_constraints}:
The largest effect of mapping only 59 source neurons to a on-wafer routing source is visible for a neuron size of 16 and low connection probabilities.
Now, the maximum number of source neurons per HICANN is reduced to $\frac{59}{64}$ of the non-constrained case.
Interestingly, for neuron sizes of 8 and 4 there is barely any effect on the expected number of source neurons.
Yet, there synapse utilization is significantly reduced especially for higher connection probabilities, as shown on the right of the figure for neuron size 8.
Note that here the number of target neurons per HICANN was reduced to 59 respectively 118 for neuron sizes 8 and 4, as this represents a more realistic mapping scenario avoiding the need for additional on-wafer routing buses for outgoing connections.

The \emph{synapse driver chain length} limitation only affects neuron sizes 8 and 4:
Here, the cases are marked by a dotted line, where the probability that more than 5 synapse drivers are needed per routing source is larger than \SI{80}{\%}.

\paragraph{Maximum recurrent networks}
\begin{figure}
	\centering
	\includegraphics{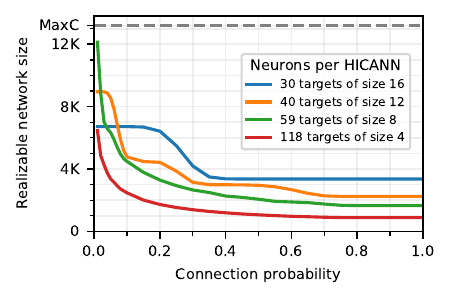}
	\caption{
		\textbf{Maximum recurrent uniform random networks.}
    \small
		The figure shows the expected maximum network size of uniform random
		networks that can be implemented on one wafer for different connection
		probabilities and mapping scenarios.
		We assume that all inter-HICANN connections can be realized during the
		on-wafer routing step (\Cref{sec:mapping}).
		For this study, the number of source neurons per routing source is
		consistent with the number neurons sending on one routing bus and
		depends on the mapping scenario.
		In contrast, previous studies (e.g., \cref{fig:sources_vs_pconn} and \cref{fig:sources_vs_pconn_comp_constraints})
		always use the same number (64 or 59) of source neurons per routing
		source.
		\label{fig:max_sized_recurrent_network}
	}
\end{figure}
So far, we have analyzed feed-forward networks on one HICANN.
Next, we ask \emph{what are the largest recurrent uniform random networks that can be realized on the wafer?}.
For this we assume a \emph{consistent mapping of neurons per routing source} for both input and target neurons on the HICANN.
For example, for a neuron size of 16, we map only 30 neurons per HICANN (and not 32) such that only two 2-MSB patterns per routing source are needed.
We now consider four different mapping setups which are specified in \cref{tab:mapping-recurrent} depending on the neuron size.
Note that we added a setup with 40 neurons of size 12.
The setups may use up to 112 or 224 HICANNs depending on used outgoing routing sources per HICANN.
We assume that all routes between HICANNs can be established in the \emph{wafer routing} step (\Cref{sec:mapping}).

The maximum realizable recurrent networks depending on connection probability are shown in \cref{fig:max_sized_recurrent_network}:
For $p<0.02$ a neuron size of 8 allows to realize networks with over \num{12000} neurons.
For probabilities between $0.02$ and $0.08$ the largest networks can be implemented with neuron size 12.
Instead, neuron size 16 yields the best results for connection probabilities $\ge 0.08$.
In the latter case, the network size varies between \num{3360} and \num{6720}, which corresponds to the cases of assigning 2 respectively 4 drivers per routing source.

\begin{table}[]
    \centering
    \begin{tabular}{l|cccc}
     \textbf{neuron size}   & \textbf{4} & \textbf{8} & \textbf{12} & \textbf{16}  \\
     \hline
     neurons per HICANN & 118 & 59 & 40 & 30 \\
     nouting sources per HICANN & 2 & 1 & 1 & 1 \\
     2-MSB patterns per routing source & 4 & 4 & 3 & 2 \\
     maximum number of HICANNs & 112 & 224 & 224 & 224 \\
     maximum number of neurons & \num{13216} & \num{13216} & \num{8960} & \num{6720} 
    \end{tabular}
    \caption{\textbf{Mapping setups to estimate maximum realizable recurrent networks in \cref{fig:max_sized_recurrent_network}.}
    For neuron size 12 there are maximum 40 neurons per HICANN and not 42 as compound neurons may not exceed the bounds of neuron blocks containing 64 neurons per priority encoder.}
    \label{tab:mapping-recurrent}
\end{table}

\subsection{Synapse loss estimation for uniform random networks}
\label{sec:results_synloss}
In the previous section we estimated the maximum sized random networks that can be completely realized on the BrainScaleS wafer.
Here, we \emph{estimate the synapse loss} for uniform random networks according to the methodology developed in \cref{sec:expected_loss}.

\paragraph{Synapse loss depending on number of inputs to HICANN} \label{sec:synloss_feedforward}

\begin{figure}
	\centering
	\includegraphics{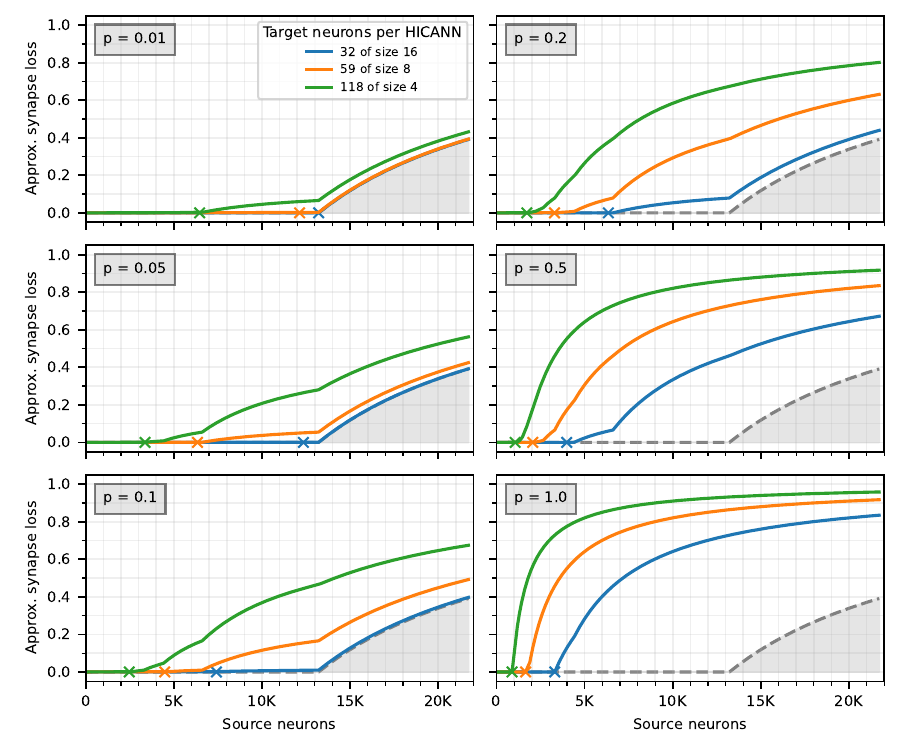}
	\caption{
		\textbf{Synapse loss approximation for feedforward uniform random networks.}
    \small
		For different connection probabilities $p$ and mapping scenarios
		(number and size of target neurons on one HICANN), the synapse loss is
		estimated depending on the number of source neurons per HICANN.
		As in \cref{fig:sources_vs_pconn_comp_constraints}, we consider the
		realistic case that only 59 of 64 addresses per routing source can be
		used for neurons (\Cref{sec:hardware-limitations}).
		The crosses mark the expected number of source neurons that can be
		implemented without synapse loss, as determined in
		\cref{fig:sources_vs_pconn_comp_constraints}.
		The gray dashed lines (highlighted by the gray area underneath) denote the synapse loss due to the limited number
		of 224 synapse drivers per HICANN:
		Once there are more than 224 routing sources, all synapses are lost for
		the fraction of routing sources exceeding this limit.
		\label{fig:expected_synloss}
	}
\end{figure}

We first consider feed-forward uniform random networks with target neurons mapped on one HICANN.
For different mapping scenarios (size and number of neurons per HICANN) we sweep the number of input neurons and connection probability $p$ and estimate the synapse loss according to \cref{eq:expected_loss} in \cref{sec:expected_loss}:
In six subplots for connection probabilities $0.01$, $0.05$, $0.1$, $0.2$, $0.5$ and $1.0$ the approximate synapse loss is plotted against the number of source neurons.
Here we only consider the realistic case with maximum 59 neurons per routing source, cf.~\cref{sec:hardware-limitations}.
As one would expect, the synapse loss increases with the connection probability and the number of source neurons, but is the smaller the larger the neuron size.
For a low connection probability of $p=0.01$ the synapse loss is close to 0 for up to 
\num{13216} source neurons for scenarios with a neuron size $\ge 8$.
After this source neuron limit, the number of routing sources exceeds the number of synapse drivers leading to an inevitable synapse loss highlighted by the gray dashed line and shaded area in the figure.
While for $p=0.01$ the synapse loss curves for neuron size 8 and 16 nearly match,
from $p=0.05$ the scenario with 32 neurons of size 16 shows significantly lower synapse loss, staying close to the synapse driver limit (dashed line) for probabilities up to $0.1$.

In \cref{fig:expected_synloss}, crosses mark the estimated maximum number of source neurons without synapse loss from the previous section (\Cref{fig:sources_vs_pconn_comp_constraints}).
In all cases, the \emph{estimated synapse loss} is $\le 0.01$ which suggests that the estimation is quite accurate.
But how do the two approaches differ? Why can the synapse loss be non-zero at the given number of input neurons? Here is the reasoning:
The maximum size estimation uses an \emph{expectation value} (a real number) of the required drivers per routing source to calculate the maximum number of source neurons (\Cref{eq:expected-drivers,eq:expected_l1sources,eq:expected_sources}), thus taking into account statistical variations.
Instead, the synapse loss estimation uses the same synapse gain \emph{distribution} $\Delta_\textrm{sorted}$ (\Cref{eq:synapse_gain_sorted}) for all routing sources and does not cope for statistical variations: When assigning $D_\textrm{ceil}$ respectively $D_\textrm{floor}$ drivers per routing source, especially the routing sources with 
$D_\textrm{ceil}$ assigned drivers may contribute to the synapse loss.
Yet, in real scenarios, the synapse gains will differ from routing source to routing source allowing for an optimized assignment.

\paragraph{Synapse loss in recurrent networks}
\begin{figure}
	\centering
	\includegraphics{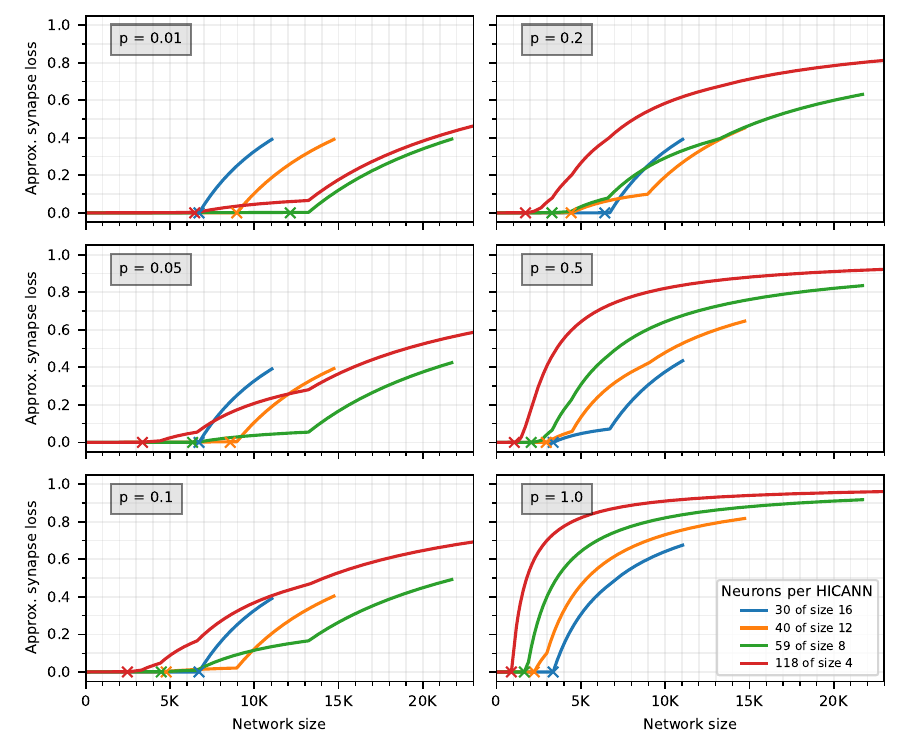}
	\caption{
		\textbf{Synapse loss approximation for recurrent networks with uniform
		random connectivity.}
    \small
		Estimated synapse loss of recurrent networks depending on the network
		size for different connection probabilities $p$ and mapping scenarios.
		We assume that all inter-HICANN connections can be realized during the
		on-wafer routing step (\Cref{sec:mapping}).
		As in \cref{fig:max_sized_recurrent_network}, the number of source
		neurons per routing source is consistent with the number neurons
		sending on one routing bus and depends on the mapping scenario.
		The crosses mark the expected network size that can be
		realized without synapse loss.
        Note that the lines end at the maximum network size per mapping scenario, which depends on the number of neurons per HICANN.
        The curves for neuron size 4 are cropped at the x-axis limit, please see \cref{fig:loss_recurrent_118} for the full plot.
		\label{fig:expected_synloss_recurrent_network}
	}
\end{figure}

Next, we look again into recurrent uniform random networks and estimate the synapse loss for different network sizes, connection probabilities and mapping scenarios.
As in the previous section, the number of source neurons per routing bus is chosen consistently with the mapping scenario.
\Cref{fig:expected_synloss_recurrent_network} shows the results in 6 subplots for different connection probabilities:
Note that the curves end at the maximum network size for each mapping scenario.
Overall, the estimated synapse loss is the smaller the larger the neuron size - until there are more routing sources than synapse drivers per HICANN which is visible as kinks at network sizes \num{6720}, \num{8960} and \num{13216} respectively for connection probabilities $<0.2$.
When exceeding these limits, some routing sources get 0 drivers assigned leading to an increasingly higher synapse loss.

The study in \cref{fig:expected_synloss_recurrent_network} can be used to pick the optimal mapping scenario for given network size and connection probability:
For networks with more than \num{21712} neurons one has to choose neuron size 4, while between \num{14720
} and \num{21712} neuron size 8  always yields the lowest synapse loss.
Interestingly, between \num{11040} and \num{14720}, for low connection probabilities ($p\le0.1$) it is best to use 59 neurons of size 8, while for higher probabilities it is best to use a neuron size of 12 (yet still with losses larger than $25\%$).
For lower network sizes, neuron sizes 12 and 16 show the lowest synapse loss, at least until the synapse driver limits (cf. kinks) are hit. 
Notably, for very low connection densities ($p=0.01$) neuron size 8 yields a close to 0 synapse loss for up to \num{13216} neurons and the lowest synapse loss until reaching the network size limit of \num{21712} neurons.
One general observation is that for a given network size not always the largest neuron size yields the lowest synapse loss, but other scenarios may perform better. Hence, one should search for the optimal mapping case by case.

\Cref{fig:expected_synloss_recurrent_network} shows the estimated maximum lossless network sizes from the previous section as crosses:
The observations are similar to those of the feed-forward networks:
For small connection probabilities ($p\leq0.2$) one can realize networks with a near-zero synapse loss whose size is significantly larger than the estimated maximum lossless network.
Instead, for denser networks, the crosses are very close to kinks where synapse loss starts to increase steeply.

We note that this synapse loss estimation for recurrent networks \emph{does not consider the potential loss of synapses during the on-wafer routing step}, when some routing sources can not be routed to their respective target HICANNs.
Yet, we can treat the results as significant for large networks by assuming that after the on-wafer routing at least 224 routing sources have been routed to each target HICANN.

\subsection{Comparison to actual mapping}
\label{sec:mapping_vs_theory}
Finally, we compare the presented synapse loss estimation with actual synapse loss after running the mapping steps in \cref{sec:mapping}.
Here, we show results using the 1st generation mapping stack, the so-called \emph{MappingTool} \cite{ehrlich2010software,bruederle2011comprehensive}. It is planned to update the results with the 2nd generation mapping stack called \emph{Marocco} \cite{muller2022operating}.

\begin{figure}
	\centering
	\includegraphics{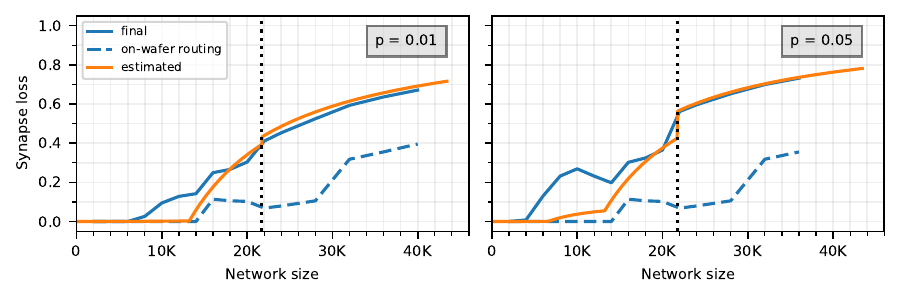}
	\caption{
		\textbf{Synapse loss of recurrent uniform random networks: mapping versus theory.}
    \small
		Relative synapse loss for recurrent networks with connection
		probability $0.01$ (left) and $0.05$ (right) at different network
		sizes.
		Labels: \textbf{final} -- synapse loss at the end of the mapping,
		\textbf{on-wafer} -- synapse loss after the on-wafer routing step,
		\textbf{estimated} -- synapse loss approximation according to
		\cref{sec:expected_loss,fig:expected_synloss_recurrent_network}.
		The vertical dashed line denotes the transition from using neuron size
		8 to 4, as in \cite[Fig.~8.8]{jeltsch2014phdthesis}.
		For $p=0.05$ networks with more than 36K neurons could not be mapped due
		to memory limitations.
        Note that the estimated curves look much smoother as they use a step size of 59 neurons, while the actual mapping is performed with larger steps. 
		\label{fig:mapping_vs_theory}
	}
\end{figure}

For this \emph{mapping versus theory study}, uniform random network with connection probabilities $0.01$ and $0.05$ are created at different network sizes and mapped to the BrainScaleS wafer using neuron sizes 4 and 8.
The synapse loss after mapping is shown in \cref{fig:mapping_vs_theory} and compared to the estimated synapse loss.
A dashed vertical line separates the neuron size used for mapping, as in
\cite[Fig.~8.8]{jeltsch2014phdthesis}.
The fraction of synapses lost after the on-wafer routing (cf.~\cref{sec:mapping}) is shown as blue dashed line.

At connection probability $p=0.01$, starting from \num{6000} neurons some synapses can not be mapped to the hardware.
At 10K neurons the synapse loss surpasses $10\%$ while the expected loss stays close to 0 until \num{13216} neurons.
This deviation can be explained as follows:
The sparseness of the synapse driver switches (\Cref{fig:wafer-routing}) makes the synapse driver assignment step (cf. ~\cref{fig:mapping}, where synapse drivers are connected to vertical buses) a hard task.
With the used connection algorithm (heuristic approach from \cite{fieres2008realizing}), for some vertical buses carrying spikes to a HICANN it can happen that there is no free synapse driver reachable via the sparse switch matrix. Hence, no synapse at all can be realized from this routing source for the given HICANN.
The effect of the sparsity is more pronounced when there are many routing sources requiring connections to synapse drivers on the HICANN.
We remark that improved results may be possible with alternative assignment algorithms as developed by \cite{jeltsch2014phdthesis}.

Between 18K and 22K the actual synapse loss approximately matches the estimated loss.
Instead, for larger networks the actual synapse loss is always slightly lower than the estimated loss.
There are two reasons for this:
First, the routing algorithm, more precisely the synapse driver requirement calculation, considers the variations of synapses per routing source and 2~MSB patterns, assigning hardware resources such that the overall loss of synapses is minimized.
Instead, the synapse loss estimation assumes identical synapse gain distributions for each routing source (as discussed in \cref{sec:expected_loss}).
Secondly, the sparseness of the synapse driver switch matrix is no longer a limiting factor, as most routing sources are connected to only one driver each.

\Cref{fig:mapping_vs_theory} on the right shows the comparison for connection probability $p=0.05$:
Notably, the actual synapse loss starts to occur from about \num{4000} neurons, increases beyond $25\%$ at \num{10000} neurons, and then falls back to $20\%$ at \num{14000} neurons.
The high loss at 10k neurons is again related to connecting vertical buses to synapse drivers which is limited by the sparsity and layout of the synapse driver switches.
At this network size, there are many routing sources with $D=2$ assigned drivers: On the one hand, if there is no free connection to a driver, many synapses are dropped. On the other hand, other routing sources may not be able to find an accessible driver due to the switch pattern.
This effect decreases when approaching \num{14000} neurons when on average each routing source per HICANN gets assigned one driver which is easier to connect.
Yet, the actual synapse loss after mapping is larger than the estimated loss between 14K and 18K neurons.
From then on, the estimation matches very well the actual synapse loss.

To summarize the results for both connection probabilities:
For small networks the synapse loss after mapping is much larger than the estimated loss, which can be attributed to challenges when connecting vertical buses to synapses drivers.
Instead, for more than 20K neurons the estimation fits quite well with the actual synapse loss.

Note that the synapse loss after the on-wafer routing in \cref{fig:mapping_vs_theory} is always much smaller  than the final synapse loss.
This suggests that the dominant factor for synapse loss in recurrent uniform random networks is the limited number of synapse drivers per HICANN, and not the inter-HICANN connectivity.
This also confirms our previous assumption in \cref{sec:expected_loss} that it is valid to ignore the on-wafer routing when estimating the synapse loss.

\section{Discussion}
\label{sec:discussion}
\subsection{Summary}
In this study we have presented theory to estimate the maximum realizable uniform random networks without synapse loss on the BrainScales wafer-scale neuromorphic system.
The approach was applied for both feed-forward and recurrent networks at different network sizes and for different mapping scenarios (neuron size respectively number of neurons per HICANN).
We further studied how specific hardware limitations (cf. \cref{sec:hardware-limitations}) decrease the maximum realizable network size.

In a second step, the theory was extended to estimate the synapse loss after mapping for given networks sizes and connection probabilities.
The results show that for a given connection probability there are much larger network sizes possible (compared to the maximum size estimation) when a synapse loss of up to \SI{1}{\%} is tolerated, especially for small $p \le 0.2$.

Next, we compared the estimated synapse loss to the real synapse loss after running the mapping software for the BrainScaleS wafer system.
Here, we used the \emph{MappingTool}, the first version of the mapping software.
For small and large recurrent network, the synapse loss estimation matches quite well the actual synapse loss after mapping.
Yet, for network sizes between \num{6000} and \num{16000} neurons, the actual synapse loss is much higher than estimated.
This can be attributed to challenges in connecting the vertical buses to synapses due to the switch sparseness between them as well as limitations of the used routing algorithm.
Improved results are expected when using the newer mapping software \cite{muller2022operating}.

\subsection{Analysis of BrainScaleS routing architecture}
We analyze the BrainScaleS architecture for uniform random networks based on our presented analytic and mapping studies:
In general, the BrainScaleS wafer-scale hardware provides a high flexibility with regard to network architectures, which is possible by combining hardware neurons to larger logical neurons with many hardware synapses. See for example \cref{fig:max_sized_recurrent_network} for the \emph{maximum recurrent uniform random networks} possible depending on neuron size and connection probability.
The \emph{synapse loss estimation} revealed that, when the number of routing sources per HICANN exceeds the number of synapse drivers, synapse loss becomes unavoidable.
While this is obvious for uniform random networks, it can also affect more structured networks as the Layer 2/3 model considered in \cite{petrovici2014characterization}.
The \emph{mapping study} showed a much higher synapse loss at intermediate network sizes which can be attributed to the sparseness of the synapse driver switch matrices (only 1 out 8 possible switches exist in a very regular pattern).
As mitigation, one could consider denser switch matrices (from which the signal quality of the 6-bit pulses may suffer) or different patterns for switch locations.
Finally, the 2b/4b decoding of 6-bit addresses adds additional complexity for the HICANN synapse routing. We compared this to a HICANN architecture where all 6-bit addresses are decoded in the synapse allowing to achieve larger lossless networks as shown in \cref{fig:sources_vs_pconn_comp_arch}.
We complement this with a synapse loss estimation in the appendix (\Cref{fig:expected_synloss_comp_arch}), showing significantly lower synapse loss for small connection probabilities ($p\le 0.2$).
The successor system BrainScales-2 \cite{pehle2022brainscales2} has adopted a 6-bit decoding in the synapse.
Note however that it has not been scaled to a wafer-scale system but instead multi-chip systems are built using FPGAs for inter-chip routing \cite{ilmberger2026brainscales2multichip}.

\subsection{Applications, limitations, related and future work}
\paragraph{Applications of the developed methods.}
Our developed methods are useful in multiple ways for users and developers of the BrainScaleS system:
\begin{itemize}
    \item One can estimate the maximum network sizes for a given connection probability.
    \item One can search for the optimal mapping scenario for a given uniform random network without running the mapping software.
    \item The synapse loss estimation can act as benchmark for the mapping software, providing an estimate of how low the synapse loss can be for a given network and mapping setup. Note however that it is possible to achieve a lower synapse loss considering the actual synaptic connections of a network instance.
    \item Our methods can be used for the design of next-generation hardware architectures for example to evaluate different address decoding schemes as in \cref{fig:sources_vs_pconn_comp_arch} and \cref{fig:expected_synloss_comp_arch}. Such studies could integrate the area overhead for more lines and SRAM cells in the synapse by reducing the number of synapse rows per HICANN.
\end{itemize}

\paragraph{Limitations}
Our approach has the following limitations:
\begin{itemize}

    \item Both maximum size estimation and synapse loss estimation do neither consider the inter-HICANN routing nor the vertical bus to synapse driver connection routing. Instead, it starts from the number of assigned drivers per routing source and assumes that there were no connections lost in the previous routing steps.
    \item Both analytic approaches do not consider all hardware features that are supported by the software and hardware: For example, we neglect considering different short-term plasticity parameters per routing source, or different synapse weight ranges that requires to use different scaling factors $\gmax$ per synapse row.
    \item The synapse loss estimation does not account for variations of synapse gains of different routing sources, as discussed in \cref{sec:synloss_feedforward}. As a result, the estimated synapse loss can be higher than the actual synapses loss if the mapping algorithms considers such variations.
    \item The methods are only shown for uniform random networks based on the binomial distribution. In principle, one can replace the binomial distribution with other discrete probability distributions. In the context of neuro-scientific modeling, one can for example use a \texttt{FixedNumberPreConnector} or \texttt{FixedTotalNumberConnector} from PyNN \cite{davison2009pynn}.
\end{itemize}

\paragraph{Related and future work}

\cite{partzsch2015network} develop principles to design hierarchical neuromorphic systems depending on the network connectivity to implement.
They derive a formalism to describe generic, multi-level connectivity structures of neuromorphic systems with synapse matrices, similar to the multi-stage address decoding in the HICANN considered here.
For different hardware configurations, they estimate the synapse loss for uniform random networks.
This is very similar to our studies in \cref{sec:results_synloss} using the methods from \cref{sec:expected_loss}.
As we do, \cite{partzsch2015network} combines the synapse loss from different levels of hierarchy.
Our work estimates the synapse loss specifically for the BrainScaleS wafer considering many features (such as combining neuron circuits) and constraints (\cref{sec:hardware-limitations}).
Beyond that, our work provides the methodology for predicting maximum realizable networks without synapse loss and the comparison to actual mapping results.
Future work could extend our methods to the generic, hierarchical connectivity structure from \cite{partzsch2015network} and could consider other than uniform-random connectivity.

Our work estimates the synapse loss for given uniform random networks on the BrainScaleS wafer.
However, taking the computational neuroscience perspective, network models should be mapped to neuromorphic hardware without any synapse loss, especially when performing parameter sweeps or trying to reproduce biological measurements.
DarwinWafer \cite{zhu2025darwinwafer} is a recent example that large-scale neuromorphic systems for neuroscience are not dead. Their paper shows differences between original and mapped connectivity of mouse and zebrafish brains, hence a measure of synapse loss.
Our approach has the potential to further improve system-level design, while compensation methods for synapse loss from \cite{petrovici2014characterization} can help to map functional brain models to neuromorphic substrates.

Beyond brain simulation, our approach can help design better multi-core neuromorphic systems for energy-efficient AI systems. Our methods are applicable to any hierarchical circuit-switched neuromorphic hardware architecture based on synapse crossbars, for both spiking and non-spiking architectures \cite{moro2022hardware,wan2022compute}. One could design architectures tailored to support certain classes of sparse or small-world connectivity \cite{billaudelle2026small}.
Later, the AI models can be optimized/trained for a given routing and hardware architecture for best utilization of hardware resources \cite{weber2025hardware}.

\section*{Acknowledgments}
The authors would like to thank Johannes Partzsch for feedback on an earlier version of this study.

The work was funded by the EC Horizon 2020 Framework Programme under grant agreements 720270 (HBP SGA1), 785907 (HBP SGA2), 945539 (HBP SGA3) and Horizon Europe grant agreements 101147319 (EBRAINS 2.0).

\bibliographystyle{apalike}
\bibliography{references}

\appendix
\section{Appendix}
\begin{figure}
    \centering
    \includegraphics[width=0.5\linewidth]{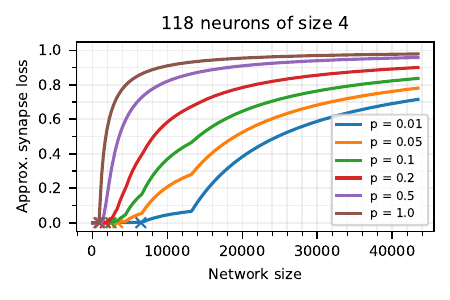}
    \caption{
        \textbf{Synapse loss approximation for recurrent networks with uniform random connectivity with 118 neurons per HICANN.}
        \small
		Estimated synapse loss of recurrent networks depending on the network
		size for different connection probabilities $p$ for the mapping scenario with 118 neurons per HICANN of size 4.
        In comparison to \cref{fig:expected_synloss_recurrent_network}, this plot shows the curves for different $p$ in one plot and up to the maximum network size for this mapping scenario.
    }
    \label{fig:loss_recurrent_118}
\end{figure}

\begin{figure}
	\centering
	\includegraphics{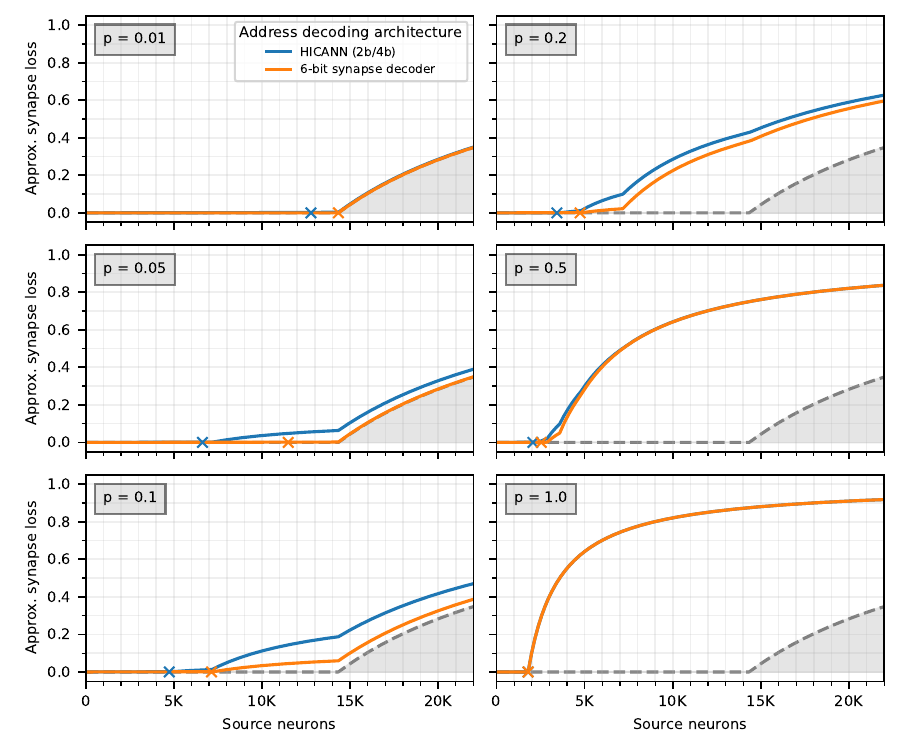}
	\caption{
		\textbf{Synapse loss approximation - Comparison of address decoding architectures.}
    \small
        We compare the estimated synapse loss for the address decoding of the HICANN (2 bits per half synapse row, 4 bits per synapse) with a 6-bit decoder in the synapse like:
		For different connection probabilities $p$ and a fixed mapping scenario (64 neurons of size 8), the synapse loss is
		estimated depending on the number of source neurons per HICANN.
        Here, we assume 64 addresses per routing source and neglect hardware limitations (\Cref{sec:hardware-limitations}).
		The crosses mark the expected number of source neurons that can be
		implemented without synapse loss, as determined in
		\cref{fig:sources_vs_pconn_comp_arch}.
		The gray dashed lines (highlighted by the gray area underneath) denote the synapse loss due to the limited number
		of 224 synapse drivers per HICANN:
		Once there are more than 224 routing sources, all synapses are lost for
		the fraction of routing sources exceeding this limit.
		\label{fig:expected_synloss_comp_arch}
	}
\end{figure}
\end{document}